 \documentclass[nohyper,11pt,letterpaper]{JHEP3}
 \usepackage{epsfig}

 \newcommand{\beq}[1]{\begin{equation}\label{#1}}
 \newcommand{\eeq}{\end{equation}}
 \newcommand{\bear}[1]{\begin{eqnarray}\label{#1}}
 \newcommand{\ear}{\end{eqnarray}}
 \newcommand{\nn}{\nonumber}

 \catcode`\@=11 \@addtoreset{equation}{section}\catcode`\@=12

 \newcommand{\R}{ \mathbb{R} }
 
 \newcommand{\e}{ \mbox{\rm e} }
 \newcommand{\eps}{ \varepsilon }
 \newcommand{\p}{\partial}
 \newcommand{\btd}{\bigtriangledown}
 \newcommand{\btu}{\bigtriangleup}

 \author{V. D. Ivashchuk and V.N. Melnikov\\
    Center for Gravitation and Fundamental Metrology,\\
    VNIIMS and Institute of Gravitation and Cosmology \\
    Peoples' Friendship University of Russia, \\
 46 Ozernaya Street, G-361, Moscow, 119361, Russia \\
    E-mail: \email{rusgs@phys.msu.ru}
 }
 \author{D. Singleton\\
 Physics Department, CSU Fresnso, \\
 Fresno, CA 93740-8031, USA \\
    E-mail: \email{dougs@csufresno.edu}
 }

\abstract{ In this paper we generalize electric $S$-brane
solutions with maximal number of branes. Previously for the action
containing $D$-dimensional gravity, a scalar field and
antisymmetric $(p+2)$-form we found composite, electric $S$-brane
solutions with all non-zero ``charge'' densities which obeyed
self-duality or anti-self-duality relations. These solutions
occurred when
$D = 4m+1 = 5, 9, 13, ...$ and  $p = 2m-1 = 1, 3, 5, ...$. 
Here we generalize these solutions to the
case when the spatial $4m$-dimensional  submanifold is Ricci-flat
rather than simply Euclidean-flat and the charge density form is a
parallel self-dual or anti-self-dual form of rank $2m$. Also
generalizations are found for the case when there is an extra
``internal'' Ricci-flat manifold not covered by the $S$-branes. 
In the case when one allows a phantom scalar field a subset of these 
solutions lead to accelerated expansion of this extra spatial factor 
space not covered by the $S$-branes while the other spatial factor space 
of dimension $4m$ contracts. Some of these $S$-brane solutions also 
provide specific examples of solutions of type $IIA$ supergravity.
}

\title{Electric $S$-brane solutions with parallel forms on Ricci-flat 
       factor space}

\begin{document}

 %%%%%%%%%%%%%%%%%%%%%%%%%%%%%%%%%%%%%%%%%%%%%%%%%%%%%%%%%%%%%%%%%%%%%%%%
 \section{\bf Introduction}
 \setcounter{equation}{0}
 %%%%%%%%%%%%%%%%%%%%%%%%%%%%%%%%%%%%%%%%%%%%%%%%%%%%%%%%%%%%%%%%%%%%%%%%

The important role played by the relationship between the charge
densities of $D$-branes in string theory was made clear in
Polchinski's work on $D$-branes \cite{Polchinsky}. 
$K$-theory (see, \cite{MM,Witten,Gukov} and references therein) 
gives a general mathematical framework for describing such
relationships between charge densities of branes.

Other types of relations between brane charge densities may follow
from the fields equations when certain gravitational background
solutions are considered. For example, in reference \cite{IS}
composite electric $S$-brane solutions were studied, and it was
found that constraints coming from the fields equations (in
particular the requirement that the off-diagonal components of the
Einstein tensor vanish) fixed the relationships between the
charges densities of the various $S$-branes. For more on $S$-brane
solutions see \cite{IS}-\cite{Ierice} and references, therein.
Specifically in $5$-dimensions with a $3$-form field and a scalar field
the non-zero charge densities of the six electric branes obeyed
the following relations
    \bear{b1}
    Q_{12} = \mp Q_{34}, \quad Q_{13} = \pm Q_{24}, \quad
     Q_{14} = \mp Q_{23}.
     \ear
   or, equivalently,
    \beq{b2}
    Q_{i j} =  \pm \frac{1}{2} \eps _{i j k l}
      Q^{k l} = \pm (* Q)_{i j} 
     \eeq
and where $Q_{i j}= - Q_{j i}$. Thus, the charge density form is 
self-dual or anti-self-dual.  
When all $Q_{i j}$ are non-zero, the configuration
from (\ref{b2}) is the only possible one that follows from the
non-diagonal part of Hilbert-Einstein equations.

One interesting feature of the $D=5$ electric $S$-brane solutions
with the $3$-form and charge densities satisfying (\ref{b1}) was that
when the scalar field was absent one was able to avoid the BKL
type \cite{BKL} oscillating behavior as one approached the initial
singularity \cite{IMS}. BKL type oscillations are asymptotical
never ending oscillations of the scale factors and the Kasner parameters
which characterize these solutions. This oscillatory behavior can be 
described graphically in terms of the billiard picture. 
The billiard approach for $D=4$
Bianchi-IX model  was introduced by Chitre \cite{Chitre}. The
multidimensional case was studied (for block-diagonal metrics
on product of Einstein spaces) in \cite{IKM1}-\cite{IMb1}. A
review of the billiard approach can be found in \cite{DHN}. 
In terms of the billiard  description the solution from \cite{IMS} 
corresponds to a non-moving (``frozen") points in the billiard. 

In this article we obtain generalizations of the solutions
investigated in \cite{IS}. The generalizations we consider are:
(i) allowing the spatial factor space covered by the $S$-branes to
be Ricci-flat rather than Euclidean flat; (ii) considering some of
the extra Ricci-flat factor space not covered by the $S$-branes.
We briefly mention possible cosmological applications of some
subset of these solutions and show that some of these solutions give 
examples of type $IIA$ supergravity solutions.

 %%%%%%%%%%%%%%%%%%%%%%%%%%%%%%%%%%%%%%%%%%%%%%%%%%%%%%%%%%%%%%%%%%%%%%%%
 \section{\bf S-brane solutions on product of flat spaces}
 \setcounter{equation}{0}
 %%%%%%%%%%%%%%%%%%%%%%%%%%%%%%%%%%%%%%%%%%%%%%%%%%%%%%%%%%%%%%%%%%%%%%%%

Here as in \cite{IS} we consider the model governed by the action
   \beq{2.1i}
    S =
       \int_{M} d^{D}z \sqrt{|g|} \left[ {R}[g] -
       g^{MN} \partial_{M} \varphi \partial_{N} \varphi
    -  \frac{1}{q!} \exp( 2 \lambda \varphi ) F^2 \right],
   \eeq
where $g = g_{MN} dz^{M} \otimes dz^{N}$ is the metric,
$\varphi$   is a  scalar field, $\lambda \in  \R$ is a
constant dilatonic coupling and
$$
   F =  dA =
   \frac{1}{q!} F_{M_1 \ldots M_{q}}
   dz^{M_1} \wedge \ldots \wedge dz^{M_{q}},
$$
is a $q$-form, $q =  p +2 \geq 1$, on a $D$-dimensional manifold $M$.
In (\ref{2.1i}) we denote $|g| = |\det (g_{MN})|$, and
$F^2 =F_{M_1 \ldots M_{q}} F_{N_1 \ldots N_{q}}
g^{M_1 N_1} \ldots g^{M_{q} N_{q}}.$

The equations of motion corresponding to  (\ref{2.1i}) are
   \bear{2.4i}
   R^M_N - \frac{1}{2} \delta^M_N R  =   T^M_N,
   \\
   \label{2.5i}
   {\btu}[g] \varphi -  \frac{\lambda}{q!}
    e^{2 \lambda \varphi} F^2 = 0,
   \\
   \label{2.6i}
   \nabla_{M_1}[g] (e^{2 \lambda \varphi}
    F^{M_1 \ldots M_{q}})  =  0.
   \ear
In (\ref{2.5i}) and (\ref{2.6i}), ${\btu}[g]$ and ${\btd}[g]$ are
Laplace-Beltrami and covariant derivative operators corresponding
to  $g$. Equations (\ref{2.4i}), (\ref{2.5i}) and (\ref{2.6i})
are,  respectively, the multidimensional Einstein-Hilbert
equations, the ``Klein-Fock-Gordon" equation for the scalar field
and the ``Maxwell" equations for the $q$-form.

The source terms in (\ref{2.4i}) can be split up as
   \bear{2.7i}
   T^M_N =   T^M_N [\varphi,g]
   + e^{2 \lambda \varphi} T^M_N[F,g],
   \ear
   with
   \bear{2.8i}
   T^M_N[\varphi,g] =
   \p^{M} \varphi \p_{N} \varphi -
   \frac{1}{2} \delta^M_N \p_{P} \varphi \p^{P} \varphi,
   \\
   T^M_N [F,g] = \frac{1}{q!} \left[ - \frac{1}{2} \delta^M_N F^2
   + q  F_{N M_2 \ldots M_{q}} F^{M M_2 \ldots M_q}\right] ,
   \label{2.9i}
   \ear
which are the stress-energy tensor of the scalar field and $q$-form,
respectively.

In \cite{IS} composite, electric $S$-brane  solutions were found to the 
equations (\ref{2.4i}), (\ref{2.5i}), (\ref{2.6i}) which were maximal in 
the sense that all the charge densities of the $S$-branes were non-zero. 
These solutions occurred in spacetimes of dimension $D= n+1 = 4m + 1= 5, 
 9, 13, \dots$ and with form field having $p = 2m - 1= 1, 3, 5, \dots$.  
In addition these solutions had non-exceptional dilatonic coupling 
     \beq{lambda}
     \lambda ^2 \neq  \frac{n}{4(n-1)} \equiv \lambda^2_0,
     \eeq
and were defined on  the manifold
   \beq{2.10g}
    M = (t_{-}, t_{+})  \times \R^{n}
   \eeq
where $\R^n$ is a flat Euclidean space.  From \cite{IS} these 
$S$-brane solutions had the explicit form
     \bear{4.pa}
      ds^2
      &=& - e^{2n \phi(t)} dt^2 + e^{2 \phi(t)}
     \sum _{i=1}^n (dy^i)^2 \\
     \label{4.pb}
     \varphi &=& \frac{n}{4(1-n)K} \left( C_2 t + C_1 \right)
               - \frac{\lambda f(t)}{K}, \\
     \label{4.pc}
     F &=& e^{2 f(t)} dt \wedge Q,   \\
     \label{4.pq}
      Q &=&  \frac{1}{(p+1)!} Q_{i_0  \dots i_p}
                           dy^{i_0}  \wedge \dots  \wedge dy^{i_p}.
     \ear
The functions $f(t)$ and $\phi (t)$ are given by
        \beq{4.i}
        f (t) = - \ln \left[|z(t) ||K Q^2|^{1/2} \right] \qquad {\rm and}
 \qquad  \phi (t)  = \frac{1}{2 (1- n) K}  
 \left[ \lambda (C_2 t + C_1) + f(t) \right],
     \eeq
where $K$ and $Q^2$ are given by
     \beq{4.k}
          K \equiv  \lambda ^2 - \frac{n}{4(n-1)} \qquad {\rm and} \qquad
          Q^2 \equiv \frac{1}{(p+ 1)!}
       \sum_{i_0,  \dots, i_p} Q_{i_0  \dots i_p}^2 > 0,
      \eeq
and $C_2 , C_1$ are integration constants. The function $z(t)$, takes one
of the four following forms depending on the value of $K$ and another integration 
constant, $C$
     \bear{4.ja}
     z(t) &=& \frac{1}{\sqrt{C}} \sinh \left[ (t-t_0) \sqrt{C} \right],
           \qquad K<0 , ~C>0; \\
                          \label{4.jb}
     &=& \frac{1}{\sqrt{-C}} \sin \left[ (t-t_0) \sqrt{-C} \right],
           \qquad K<0 , ~C<0; \\
                           \label{4.jc}
     &=& t-t_0, \qquad \qquad \qquad \qquad \qquad K<0, ~C=0; \\
                            \label{4.jd}
      &=& \frac{1}{\sqrt{C}} \cosh \left[ (t-t_0) \sqrt{C} \right],
     \qquad K>0 , ~C>0,
     \ear
Here due to  the zero energy constraint \cite{IS}
 $$C \equiv \frac{n }{4(n-1)} (C_2)^2 \geq 0 .$$ 
Under this condition we exclude the solution (\ref{4.jb}) above. Later 
when we discuss the general case we will consider this solution.

The $Q_{i_0  \dots i_p}$ are constant components of the charge density form $Q$
which obey self-duality or anti-self-duality relations \cite{IS}, i.e.
$$
      Q_{i_0  \dots i_p} =
     \pm \frac{1}{(p+1)!} \eps _{i_0  \dots i_p j_0 \dots j_p}
      Q^{j_0 \dots j_p} = \pm (* Q)_{i_0  \dots i_p}.
$$
This solution describes a collection of 
$k \leq \frac{(4m)!}{(2m)! ^2}$ electric $Sp$-branes with
non-zero charge densities $Q_{i_0  \dots i_p} \neq 0$, $i_0 <  \dots < i_p$.

The above solutions, can also be enlarged by allowing multiple
scalar fields \cite{DIM}.

 %%%%%%%%%%%%%%%%%%%%%%%%%%%%%%%%%%%%%%%%%%%%%%%%%%%%%%%%%%%%
 \section{\bf Generalization to Ricci-flat factor space}
 %%%%%%%%%%%%%%%%%%%%%%%%%%%%%%%%%%%%%%%%%%%%%%%%%%%%%%%%%%%%

We now show that one can generalize the solution from the previous section to the case
when the manifold (\ref{2.10g}) is replaced by the manifold
    \beq{3.10g}
    M = (t_{-}, t_{+})  \times N,
   \eeq
where $N$ is $n$-dimensional oriented  manifold with a Ricci-flat
metric $h = h_{ij}(y)dy^i \otimes dy^j$ of Euclidean
signature. As before we will find $S$-brane solutions  when $D=n+1
=4m+1$, $m =1,2, \dots$.
The charge density form (\ref{4.pq}) generalizes to
$$
      Q =  \frac{1}{(p+1)!} Q_{i_0  \dots i_p}(y)
              dy^{i_0}  \wedge \dots  \wedge dy^{i_p},
$$
where the components are now $y$-dependent. The solutions will
again occur when the rank of $Q$ is $2m$ and the rank of the form
field is $2m+1$.

The form $Q$ is required to be parallel, i.e. covariantly constant, with
respect to $h$
     \beq{3.12a}
      (i) \quad \btd[h] Q = 0,
     \eeq
and also self-dual or anti-self-dual
       \beq{3.12b}
      (ii) \quad  Q = \pm * Q.
     \eeq
Here $* = *[h]$ is Hodge operator for the metric $h$.
It follows from (i) that
  \beq{3.12ca}
        \quad Q^2 \equiv \frac{1}{(p+ 1)!}
       h^{i_0 j_0} \dots h^{i_p j_p}
       Q_{i_0  \dots i_p} Q_{j_0  \dots j_p},
  \eeq
is constant. This in turn implies that $Q^2 > 0$, since $Q$ is
non-zero and the metric $h$ has Euclidean signature.

For non-exceptional value of dilatonic coupling (\ref{lambda}) we
will show that the form of the solution given in (\ref{4.pa}) -
(\ref{4.jd}) carries over to the present case simply by replacing
Euclidean flat metric  by the Ricci-flat one, i.e.
     \bear{5.pa}
      ds^2
      &=& -e^{2n \phi(t)} dt^2 + e^{2 \phi(t)}
      h_{ij}(y)dy^i dy^j, \\
     \label{5.pb}
     \varphi &=& \frac{n}{4(1-n)K} \left( C_2 t + C_1 \right)
               - \frac{\lambda f(t)}{K}, \\
     \label{5.pc}
     F &=& e^{2 f(t)} dt \wedge Q,
     \ear
provides a solution to the field equations (\ref{2.4i}) -(\ref{2.6i}) for the manifold
(\ref{3.10g}). As before the functions $f(t)$ and $\phi (t)$ are given by  (\ref{4.i}),
$K$ is defined in  (\ref{4.k}), $C_2 , C_1$ are integration constants, $z(t)$ is
given by (\ref{4.ja})-(\ref{4.jd}), and $C = \frac{n}{4(n-1)} (C_2)^2$.

The scalar field in (\ref{5.pb}) still solves (\ref{2.5i}) since
$\varphi$ only depends on $t$ and, for the block diagonal metric
considered here, going from Euclidean flat to Ricci-flat
metric, $h$, only modifies
the spatial parts of the Laplace-Beltrami operator. The form field
(\ref{5.pc}) also still solves the  ``Maxwell'' equation,  (\ref{2.6i}). The time
dependent part of $F$ still works in (\ref{2.6i}) for the same reason as
for the scalar field. The one possible deviation could come from the covariant
derivative operator acting on $Q$, but from (\ref{3.12a}) one sees this does not
give any additional contribution.

To conclude we need to show that the metric given by the line
element (\ref{5.pa}) still satisfies the field equation
(\ref{2.4i}). To verify the Hilbert-Einstein
eqs. (\ref{2.4i}), we first show that
   \beq{5.d}
     T[F, g]_i^{~j} = 0,
   \eeq
for all $i,j = 1, \dots, n$. Physically this means that when
$\lambda = 0$ the form field contributes as dust matter.

In what follows we use the following notation
    \beq{5.f}
     C_i ^{~j} =
     \sum_{i_1, \dots, i_p =1}^{n}
      Q_{i i_1 \dots i_p}  Q^{j i_1 \dots i_p}
     \eeq
For $i \neq j$, $T[F,g]_i^{~j}$ is proportional to $C_i ^{~j}$.

First, prove the relation (\ref{5.d}) for $i \neq j$. Due to the
(anti-) self-duality of $Q$ we get
     \beq{5.ff}
      C_i ^{~j}
     = \sqrt{|h|}
     \sum_{i_1, \dots, i_p =1}^{n}  \sum_{j_0, \dots, j_p =1}^{n}
     \pm \frac{1}{(p+1)!} \eps _{i i_1 \dots i_p j_0 \dots j_p}
     Q^{j_0 \dots j_p}   Q^{j i_1 \dots i_p}.
    \eeq
This can be further rewritten as
     \bear{5.g}
    C_i ^{~j} &=&   \sqrt{|h|}
   \sum_{i_1, \dots, i_p =1}^{n}  \sum_{j_1, \dots, j_p =1}^{n} \pm
   \frac{1}{p!}\eps _{i i_1 \dots i_p j j_1 \dots j_p}
    Q^{j j_1 \dots j_p}   Q^{j i_1 \dots i_p}
   \nonumber \\ &=& \sqrt{|h|}
   \sum_{i_1, \dots, i_p =1}^{n}  \sum_{j_1, \dots, j_p =1}^{n}
   \pm \frac{(-1)^p}{p!}\eps _{i j_1 \dots j_p  j i_1 \dots i_p}
   Q^{j i_1 \dots i_p}  Q^{j j_1 \dots j_p}
   \nonumber \\
   &=& (-1)^p  C_i ^{~j}.
      \ear
Note that $j$ is not summed over in the two sums above, and we have explicitly
written out the sums that are performed.
In going from  (\ref{5.f}) to the first
line of  (\ref{5.g}) we have carried out $p + 1$ identical
sums with:  $j_0 = j$,  $j_1 = j$, ..., $j_p = j$, respectively.
From (\ref{5.g}) one finds that  for odd $p = 2m - 1$
$$ C_i ^{~j} = - C_i ^{~j} \Rightarrow
C_i ^{~j}=0, \quad i \neq j ~,$$
and, hence, relation (\ref{5.d}) is valid for $i \neq j$.

Next we prove relation (\ref{5.d}) for $i = j$, i.e.
    \beq{5.9}
      T^i_i [F,g] = 0,
    \eeq
(no summation in $i$) for all $i = 1, \dots, n$.

It follows from (\ref{2.9i}) and (\ref{4.pc})
    \beq{5.9i}
      T^i_i [F,g] = B(t) \left[ - \frac{1}{2} \sum_{k =1}^n C^k_k + (q -1) C^i_i \right],
    \eeq
for all $i = 1, \dots, n$ where $B(t)$ is function of $t$. The
matrix  $(T^i_j [F,g])$ is traceless

$$ \sum_{k =1}^n  T^k_k [F,g] = 0, $$
since $n = 2 (p + 1) = 2 (q - 1)$. To prove  (\ref{5.9}) it
sufficient to verify that $T^1_1  = \dots = T^n_n$, or,
equivalently, $C^1_{~1}  = \dots = C^n_{~n}$.

Next we show without restriction of generality
that $C^1_1  =  C^2_2$. Indeed, using (\ref{5.ff})
we get (the summation over repeated indices is understood)
     \bear{5.12}
    C_1^{1} &=&  \pm   
    \frac{\sqrt{|h|}}{(p+1)!}\eps _{1 i_1 \dots i_p j_0  \dots j_p}
            Q^{j_0 j_1 \dots j_p}   Q^{1 i_1 \dots i_p}
    \\ \nonumber
        &=&  \pm  
        \frac{\sqrt{|h|}}{(p+1)!} 
        \left[ p \eps _{1 2 i_2 \dots i_p j_0 j_1 \dots j_p}
           Q^{j_0 j_1 \dots j_p}   Q^{1 2 i_2 \dots i_p}
         +  (p + 1) \eps _{1 i_1 \dots i_p  2 j_1 \dots j_p}
                    Q^{ 2 j_1 \dots j_p}   Q^{1 i_1 \dots i_p} \right]
      \\ \nonumber
        &=&  \pm  \frac{\sqrt{|h|}}{(p+1)!}
        \left[ p \eps _{2 1 i_2 \dots i_p j_0 \dots j_p}
        Q^{j_0 j_1 \dots j_p}   Q^{ 2 1 i_2 \dots i_p}
        +  (p + 1) \eps _{2 j_1 \dots j_p  1 i_1 \dots i_p}
           Q^{1 i_1 \dots i_p} Q^{ 2 j_1 \dots j_p} \right]
     \\ \nonumber
         &=&  \pm  \frac{\sqrt{|h|}}{(p+1)!}
         \eps _{2 i_1 \dots i_p j_0 j_1 \dots j_p}
         Q^{j_0 j_1 \dots j_p}   Q^{2 i_1 \dots i_p}
      \\ \nonumber
         &=&  C_2^{2}.
        \ear
Here we used the fact that $p+1 = 2m$ is even.

This completes the demonstration of (\ref{5.d}). Now we can
verify the Hilbert-Einstein equations (\ref{2.4i}) for the Ricci-flat
case of (\ref{5.pa})-(\ref{5.pc}). The Hilbert-Einstein equations 
are satisfied for the non-diagonal components ($M \neq
N$), since the Einstein tensor on the left hand side of (\ref{2.4i}), for the metric
(\ref{5.pa}), is diagonal (see the appendix in \cite{IMtop})
and the stress-energy tensor (\ref{2.7i}) $T^M_N$ is also
diagonal due to eqs. (\ref{5.d}) and because the scalar
field only has a temporal dependence -- $\varphi = \varphi(t)$. 
Next, because of the Ricci-flatness of $h$ and (\ref{5.d}) 
the diagonal part of the Hilbert-Einstein 
equations (\ref{2.4i}) give the same ordinary differential equations for 
the metric warp factor, $\phi(t)$, (with
constant parameter $Q^2$ from  (\ref{3.12ca})) as in the Euclidean flat
case. Thus the solution for $\phi (t)$ is again given by 
(\ref{4.i}) \cite{IS}.

This shows that the solutions given in (\ref{5.pa}) -
(\ref{5.pc}) satisfy the field equations (\ref{2.4i}) -
(\ref{2.6i}) when the  metric, $h$, is generalized
from Euclidean to Ricci-flat. As
in the Euclidean flat case the solutions correspond to  
$D = 4m +1$ and $p=2m-1$.

  \subsection{Generalization to extra Ricci-flat space not covered by 
              $S$-brane}

In this subsection we give a generalization of the solution from
the previous section when the manifold of (\ref{3.10g}) is replaced by
$$
    M = (t_{-}, t_{+})  \times N \times N_1 ,
$$
where $N_1$ is a Ricci-flat manifold with the metric $h^1$, of dimension 
$d_1$ which is not covered by the $S$-branes.  

For manifold above we now find
     \bear{4A.pai}
   ds^2 &=&  e^{\frac{4m f(t)}{K(2-D)}}
   \left[ - e^{2c t + 2 \bar c} dt^2
     + e^{(\frac{f(t)}{K} + 2c^0 t + 2 \bar c^0)} h_{ij} (y) dy^i dy^j
     +   e^{2c^1 t+2 \bar{c}^1} ds^2_1 \right] ,
      \\  \label{4A.ppi}
    \varphi &=& - \frac{ \lambda}{K} f(t)
     +  c_{\varphi} t + \bar c_{\varphi} ,  \\
     \label{4A.pci}
     F &=& e^{2 f(t)} dt \wedge Q ,
         \ear
where  
 $ds^2_1 = h^1_{m n}(z_1) dz_1^{m} dz_1^{n}$ is line element
corresponding to  the metric $h^1$,  $f(t)$ is given by (\ref{4.i}) and
 (\ref{4.ja}), (\ref{4.jb}), (\ref{4.jc}), (\ref{4.jd}). The constants 
 $c$, $\bar c$ and $K$ are given by 
 
 $$ c  = 4mc^0 + d_1 c^1 \qquad \bar c  
     = 4m \bar c^0 +  d_1 \bar c^1 $$ and $$ K  =   \lambda^2 + m + 
     \frac{4m^2}{2-D} \neq 0.  $$ 
     
 where now $D=4m +1 +d_1$. 
The integration constants obey the following relations:
\bear{x}
     &&C K^{-1} + (c_{\varphi})^2 + 4 m (c^0)^2 + (c^1)^2 d_1
     - (4 m c^0  + c^1 d_1)^2 = 0, \label{x.1} \\
     &&2m c^0 = \lambda c_{\varphi}, \qquad \qquad
     2m \bar c^0 = \lambda \bar c_{\varphi} \label{x.2}.
\ear
When internal space $N_1$ is omitted we recover the solution from
the previous subsection with the following identifications between
constants:
$$
     c_{\varphi}  = \frac{C_2 n}{4(1-n)K}, \qquad
     \bar c_{\varphi}  = \frac{C_1 n}{4 (1-n)K}.
$$

For a flat Euclidean space -- $N = \R^{4m}$ -- the solution
presented above can be obtained as a special (one-block) case of 
the so-called block-orthogonal, composite $S$-brane solutions given in 
\cite{IMtop,Isbr}.  Block-orthogonal solutions first appeared in \cite{Br} 
for configurations having one factor space of non-zero 
curvature and were subsequently generalized in \cite{IMtop,Isbr,IMJ2,IMJ1}.

For a Ricci-flat space $N$ the solution under consideration may be verified
just along the lines of the previous section when the
``internal" space $N_1$ was absent. Here we get the same ordinary 
differential equations for the ansatz functions which depend only on time 
as in the case of flat $N$.   

We note that previously (when $N_1$ was absent)  
solution (\ref{4.jb}) was excluded, by the zero energy constraint.            
In the present case the new zero energy constraint (\ref{x.1})
allows solution (\ref{4.jb}).

The preceding analysis can simply be generalized to the case when there are 
several Ricci-flat spaces not covered by $S$-branes, i.e.  when $M = 
(t_{-}, t_{+})  \times N \times N_1 \times N_2 \times ...$. This was
done in \cite{GIRS} for the case when the scalar field was absent.

\subsection{Solution with acceleration} 

If one considers the above solutions in the simple case when the
integration constants vanish (i.e. $C = c_{\varphi} = c^0 = c^1 =
 0$) one finds the physically interesting solution with accelerated
expansion for $N_1$. Under these conditions the solution takes the
form given by (\ref{4.jc}) and the explicit form of $f(t)$ is
 $f(t) = -\ln \left[ |t-t_0| |K Q^2|^{1/2} \right]$. For this
solution $K < 0$ or $\lambda^2 <  m \frac{1 - d_1}{D-2}$. This is
possible when $\lambda$ is pure
imaginary. This implies that the scalar field, after the
redefinition $\varphi \rightarrow  i \varphi$,
is a phantom field \cite{phantom}, i.e. a scalar field
with a negative kinetic energy term. 

The metric  (\ref{4A.pai}) for this case reads
     \beq{5.pai}
       ds^2 = - d \tau^2 +
       B_0 \tau^{2 \nu_0} h_{ij}(y)dy^i dy^j
       + \tau^{2 \nu_1}  B_1 d s^2_1,
     \eeq
where $\tau > 0$ is ``synchronous" time variable given by
$$
 d \tau^2  = e^{\left(\frac{4m f(t)}{K(2-D)} \right)} dt^2
$$
Solving for $\tau$ gives
$$
 \tau \propto |t-t_0|^{\frac{2m+K(D-2)}{K(D-2)}}
$$
In (\ref{5.pai}) $B_0, B_1$ are positive 
constants and exponents $\nu _0, \nu_1$ are
 $$
       \nu_0 = \frac{4m + 2 - D}{2 \Delta},
       \qquad    \nu_1 = \frac{2m}{\Delta}.
 $$
Here $\Delta  = (D-2) K + 2m$. When $\nu_1 > 1$, or $ - 2m/(D-2) <
K < 0$ we get an accelerated expansion of factor space $N_1$. This
takes place when
 $$
       -(d_1+1) m  < \lambda^2  (D-2)  < - (d_1 -1) m.
 $$
For this range of $K, \lambda$ one also finds that $\nu _0 < 0$ if
 $d_1 > 1$. Thus the Ricci-flat factor space covered by the
 $S$-branes contracts for $d_1 > 1$, while the other factor space
with the line element $ds_1 ^2$, expands with accelerated
expansion. 

As a final comment we note that the 
solutions of this section for $\lambda = 0$ are in agreement with the 
perfect fluid solutions from \cite{IM95}.

 \subsection{$II A$ supergravity solutions}

In this subsection we show how the solutions of this section
provide specific examples of supergravity solutions.
In $D=10$ $IIA$ supergravity the bosonic part of the action is given by
 \beq{6.1}
 S=\int d^{10}z\sqrt{|g|} 
 \left[ {R[g]-(\p\varphi)^2
 -\sum_{a=2}^4 \e^{2\lambda_a \varphi }F_a^2} \right]
  -\frac12\int F_4\wedge F_4\wedge A_2,
 \eeq
where $F_a=dA_{a-1}+\delta_{a4}A_1\wedge F_3$ is an $a$-form and
 \beq{6.2}
  \lambda_3=-2\lambda_4, \quad \lambda_2=3\lambda_4, \quad
 \lambda_4^2=\frac18.
 \eeq

The example we consider here  corresponds to zero forms $A_1$,
$A_3$ (and hence $F_2$ and $F_4$)  in (\ref{6.1}) . This is the
so-called NS-NS (Neveu-Schwarz) sector of the model, and the
solution from this section gives a solution to this sector of the
supergravity model. It can be seen that in this case the solution
describes a collection electric S1-branes, i.e. S-fundamental
strings (SFS). In this case we have $m=1$, and $K = 1$, and the
solution reads
 \bear{4A.paj}
     ds^2 &=&  e^{\frac{-f(t)}{2}} \left[ -e^{2c t + 2 \bar c} dt^2
     + e^{f(t) + 2c^0 t + 2 \bar c^0} h_{ij}dy^i dy^j
     + e^{2c^1 t+2 \bar{c}^1} ds^2_1 \right],
      \\  \label{4A.pbi}
     \varphi &=& - \lambda_3 f(t) +  c_{\varphi} t + \bar c_{\varphi},  \\
     \label{4A.pcj}
     F_3 &=& e^{2 f(t)} dt \wedge Q.
         \ear
Here the function $f(t)$ is given by  relation
  $$f (t) = - \ln \left[|z (t) || Q^2|^{1/2} \right]$$
with
$$z (t) = \frac{1}{\sqrt{C}} \cosh \left[ (t-t_0) \sqrt{C} \right] ~,$$
and the integration constants obey
  $$c  = 4c^0 + d_1 c^1,
     \qquad  \bar c  = 4\bar c^0 +   d_1 \bar c^1,
  $$
and
     \bear{4A.haij}
     C &=& - (c_{\varphi})^2 - 4 (c^0)^2
     -  (c^1)^2 d_1  +  (4 c^0  + c^1 d_1)^2 > 0,    \nn \\
     2 c^0 &=& \lambda_3 c_{\varphi}, \qquad \qquad
     2 \bar c^0 = \lambda_3 \bar c_{\varphi}. \nn
     \label{4A.hdij}
      \ear
The inequality $C > 0$ is an important restriction
here. It implies that $c_{\varphi}$ and the ``anisotropy'' parameter
$\delta c = c^1- c^0$, should  be small enough.
We note that the solution of
the previous section does not provide a solution for $IIB$
supergravity model. In  $IIB$ supergravity model the 5-form,
$F_5$, should be self-dual and in our solution it is not.

 %%%%%%%%%%%%%%%%%%%%%%%%%%%%%%%%%%%%%%%%%%%%%%%%%%%%%%%%%%%%%%%%%%%%%%
    \section{Conclusions and discussions}
 %%%%%%%%%%%%%%%%%%%%%%%%%%%%%%%%%%%%%%%%%%%%%%%%%%%%%%%%%%%%%%%%%%%%%%

In this paper we generalized the composite electric $Sp$-brane
solutions from \cite{IS} for $D = 4m+1 = 5, 9, 13, ...$ and   $p =
 2m-1 = 1, 3, 5, ...$ to the case when $Q$-form  of rank $2m$ is
defined on $4m$-dimensional  oriented Ricci-flat factor space $N$
of Euclidean signature. Here the form $Q$ is an arbitrary parallel
self-dual or anti-self-dual $2m$-form on $N$ with $Q^2 > 0$.  For
flat $N = \R^{4m}$ \cite{IS} the components of this form in
canonical coordinates are proportional to the charge densities of
the electric $p$-branes. In addition we generalized these
solutions to the case when there was an extra Ricci-flat factor
space not covered by the $S$-branes. These generalized solutions also 
provided examples of solutions to IIA supergravity models.
One could also extend the
solutions of this paper by allowing multiple scalar
fields as in \cite{DIM}.

For the case with a phantom field a certain subclass of 
these solutions with an extra   Ricci-flat factor space had the 
interesting feature of the extra factor space having accelerated expansion 
while the space covered by the $S$-branes contracted. 

We note that such parallel forms exist when the $4m$-dimensional
manifold, $N$, is a K{\"a}hler, Ricci-flat manifold of holonomy
group $SU(2m)$. Indeed
the $m^{th}$ wedge power of a K{\"a}hler 2-form, i.e. $\alpha =
 \Omega^m$, gives an example of non-zero parallel (i.e. covariantly
constant) form of rank $2m$. Splitting this form into a sum of
self-dual and anti-self-dual parallel forms: $\alpha = \alpha_{+}
 + \alpha_{-}$, (here $\alpha_{\pm} = \frac12 (1 \pm *)\alpha$,
where * =*[h] is the Hodge operator on $N$)   we get that either
 $\alpha_{+}$ or $\alpha_{-}$ is a non-zero parallel form. Thus, we
get an example of either self-dual or anti-self-dual parallel
 $2m$-form on a K{\"a}hler, Ricci-flat manifold of dimension $4m$.
When $N$ is a hyper-K{\"a}hler,  Ricci-flat manifold of dimension
 $4m$ with holonomy group $Sp(m)$ there are three K{\"a}hler
2-forms: $\Omega_1, \Omega_2, \Omega_3$. In this case we have more
examples of parallel forms, since any wedge product $\alpha =
 \Omega_1^{m_1} \wedge \Omega_2^{m_2} \wedge \Omega_3^{m_3}$, with
$m_1 + m_2 + m_2 = m$, is a parallel form. Finally we mention that
there exists a parallel (self-dual) 4-form on a 8-dimensional, Ricci-flat
manifold of $Spin(7)$ holonomy. See item 10.124 (Table 1) in \cite{Besse}.

 \begin{center}
  {\bf Acknowledgments}
 \end{center}

The work of V.D.I. was supported in part by grant
of College of Science and Mathematics of  California State University
 (Fresno) and by a DFG grant
 Nr. 436 RUS 113/807/0-1 and by the Russian Foundation for
Basic Researchs, grant Nr. 05-02-17478.

V.D.I. thanks colleagues from the Physics Department of
the California State University of  for their hospitality during his visit
in November-December, 2005. We also thank S. Gukov for
 informing us on certain mathematical topics of this paper.

\end{document}